\documentclass[%
 aip,
 amsmath,amssymb,
reprint,%
]{revtex4-1}
\usepackage{lineno}
\usepackage{graphicx}% Include figure files
\usepackage{dcolumn}% Align table columns on decimal point
\usepackage{bm}% bold math
\usepackage{lipsum}%to allow long equations to spa

\usepackage{setspace}
\usepackage{color}
\usepackage{soul}

\usepackage{xcolor}

\begin{document}
\title{Antiferromagnetic insulatronics: spintronics in insulating 3d metal oxides with antiferromagnetic coupling}
%Including the Authors/preliminary order 

\author{H. Meer}
\affiliation{Institute of Physics, Johannes Gutenberg University Mainz, 55099 Mainz, Germany}
\author{O. Gomonay}
\affiliation{Institute of Physics, Johannes Gutenberg University Mainz, 55099 Mainz, Germany}

\author{A. Wittmann}
\affiliation{Institute of Physics, Johannes Gutenberg University Mainz, 55099 Mainz, Germany}

\author{M. Kl{\"aui}}
\email{Klaeui@Uni-Mainz.de}
\affiliation{Institute of Physics, Johannes Gutenberg University Mainz, 55099 Mainz, Germany}
\affiliation{Centre for Quantum Spintronics, Department of Physics, Norwegian University of Science and Technology, 7034 Trondheim, Norway}

\begin{abstract}
Antiferromagnetic transition metal oxides are an established and widely studied materials system in the context of spin-based electronics, commonly used as passive elements in exchange bias-based memory devices. Currently, major interest has resurged due to the recent observation of long-distance spin transport, current-induced switching, and THz emission. As a result, insulating transition metal oxides are now considered to be attractive candidates for active elements in novel spintronic devices. Here, we discuss some of the most promising materials systems and highlight recent advances in reading and writing antiferromagnetic ordering. This article aims to provide an overview of the current research and potential future directions in the field of antiferromagnetic insulatronics.
%Transition metal oxides have long been the focus of antiferromagnetic spintronics. After the observation of antiferromagnetic domains, interest in them has resurged for application as passive elements in exchange bias-based devices. However, with the recent observations of long-distance spin transport, current-induced switching, and THz emissions, insulating transition metal oxides are now considered to be attractive candidates for active elements in spintronic devices. Here, we discuss some of the most promising materials, their spin structures and we highlight recent advances in reading and writing their magnetic ordering. This paper aims to provide an overview of the exciting field of antiferromagnetic insulatronics and its future.

\end{abstract}
%\keywords{asfasfadsfadsfadsf}
\maketitle
\newpage

%%%%%%%%%%%%%%%%%%%%%%%%%%%%%%%%%%%%%%%Outline%%%%%%%%%%%%%%%%%%%%%%%%%%%%%%%%%%%%%%%%%%%%
\tableofcontents
\newpage

%%%%%%%%%%%%%%%%%%%%%%%%%%%%%%%%%%%%%%%Manuscript%%%%%%%%%%%%%%%%%%%%%%%%%%%%%%%%%%%%%%%%%%%%
%%%%%%%%%%%%%%%%%%%%%
% Introduction
%%%%%%%%%%%%%%%%%%%%%%

\section{Introduction}
In spin-based electronics, writing, storing, and reading information relies on the electron's spin rather than its charge. Spintronic devices are commonly implemented in ferromagnets~\cite{Wol2001}. Despite major advances, real devices utilizing conventional ferromagnetic metals and spin-polarised charge currents have several drawbacks: parasitic magnetic stray fields, intrinsically low characteristic dynamic frequencies, large magnetic damping, and ohmic losses. These limit the device density, integration, and operation speed, as well as increase the power consumption.

AFMs have moved to the forefront of condensed matter physics and especially spintronics, due to their unique and favorable properties, which have recently started to be exploited~\cite{MacDonald2011a, Gomonay2014d,Gomonay2017,Baltz2018,Fukami2020}. In particular, the zero net magnetic moment makes AFMs insensitive to external stray fields, thus enhancing their stability. Furthermore, the absence of stray fields implies that there is no dipolar coupling between different areas in an AFM. If used for storage, this could lead to more than a 100-fold increase in the potential storage density~\cite{Loth2012}.
In collinear AFMs, 180$^\circ$ reversal of the magnetic ordering is not easily detectable. Therefore, in contrast to ferromagnets, where a logic "1" and a logic "0" are commonly encoded by 180$^\circ$ reversal of the magnetization, information is stored along multiple directions in AFMs.
Beyond the static properties, the exchange enhancement of the dynamics~\cite{Kittel1951} leads to eigenmode frequencies that are orders of magnitude higher compared to ferromagnets. The ultrafast dynamics holds promise for antiferromagnetic devices with THz operation speed~\cite{Wienholdt2012a,Nemec2018,Kampfrath2011}. 

The three key strategies for developing novel devices for the next-generation information and communication technologies are thus to
i) eliminate stray fields to increase the density,
ii) integrate low-damping insulators to decrease the power consumption and increase the efficiency and,
iii) employ materials with ultrafast dynamics to increase the operation speed.

Theoretically, it was predicted that pure spin currents can be generated, transported, and used in antiferromagnetic insulators for writing, reading, and transporting spin information to enable such new devices~\cite{Gomonay2014d,Jungwirth2016,MacDonald2011a}.
The experimental verification for electrical detection and control of the antiferromagentic order~\cite{Wadley2016a} has recently propelled AFMs into the limelight and paves the path towards utilizing AFMs as active components in spintronic devices.
Thus, insulating AFMs hold promise for a new generation of ultrafast low-power spin-based electronic devices \cite{Cheng2016}.

Insulating magnetic oxides have been of particular interest due to their tunable magnetic ordering, magnetic properties and their chemical stability~\cite{Banerjee2019, Bib2007, Tri2022}. While ferro- and ferrimagnetic oxide systems have been widely studied, insulating antiferromagnetic oxides have recently also gained significant interest.
Most conspicuous are 3d metal (Fe, Ni, Co, Cr, etc.) oxides that commonly exhibit antiferromagnetic order with well-defined spin structures and can be grown in high quality both as bulk crystals and thin films.
This review presents an overview of insulating antiferromagnetic 3d oxide materials and discusses recent developments in detecting and manipulating the antiferromagnetic state based on a variety of mechanisms ranging from static to ultrafast phenomena.

%%%%%%%%%%%%%%%%%%%%%
% Spin Structures
%%%%%%%%%%%%%%%%%%%%%%

%\section{Materials and their Spin structures}
%\textit{Microscopic origin – strain, exchange, anisotropy}

\begin{figure*}[ht]
    \centering
    \includegraphics[width = 0.6\textwidth]{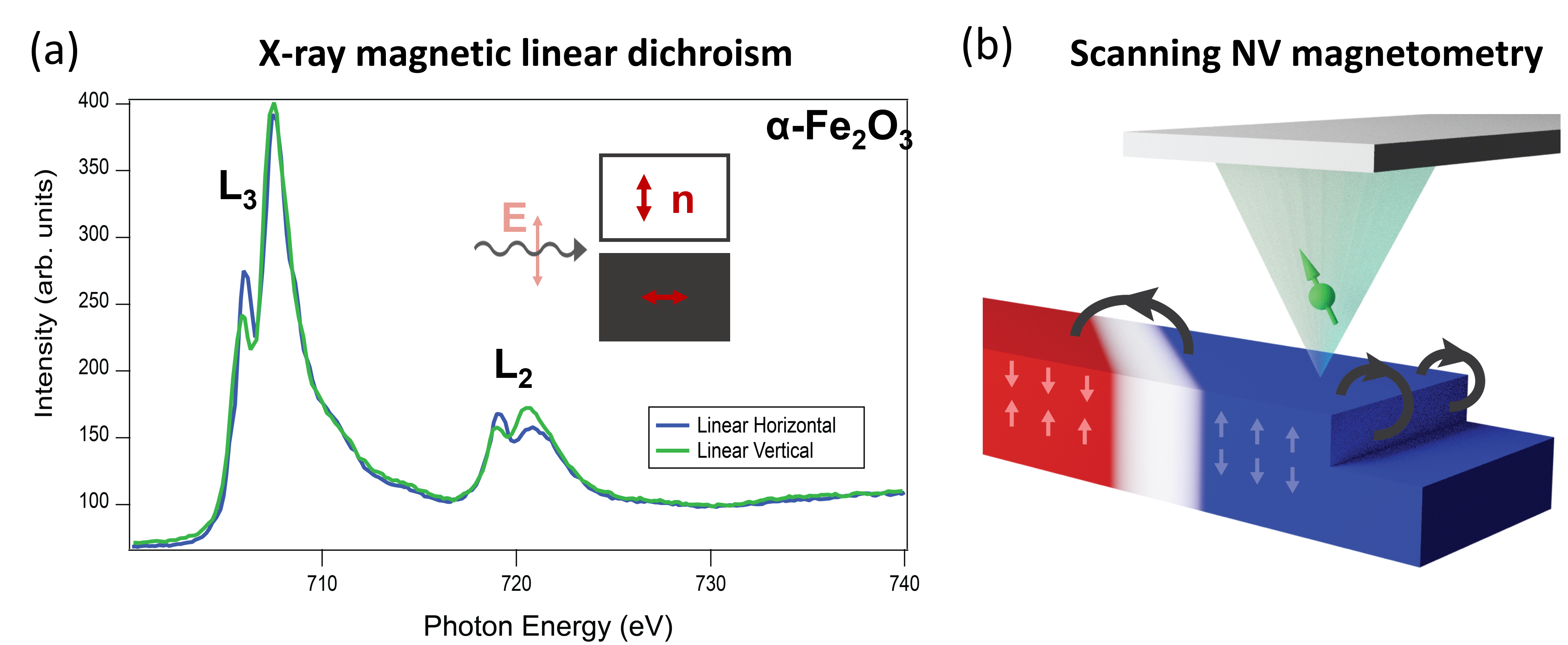}
    \caption{(a)~X-ray absorption spectra for linear horizontal (blue) and vertical (green) polarization at the Fe L$_{3,2}$ edges for $\alpha$-Fe$_2$O$_3$. Inset shows the bright (dark) contrast for parallel (perpendicular) alignment of the N\'eel vector (red) and the X-ray polarization. (b)~Illustration of imaging the stray field of AFMs at domain walls and topographic features using scanning NV magnetometry.}
    \label{fig:Imaging}
\end{figure*}
\section{Imaging methods to reveal the magnetic ordering in antiferromagnets}
A key challenge for realizing antiferromagnetic devices is the readout of the antiferromagnetic state. The absence of net magnetic moments and stray fields makes it more challenging to detect the antiferromagnetic order compared to their ferromagnetic counterparts. However, recent developments in experimental magnetic imaging techniques have led to easier access to the antiferromagnetic domain structure with increased spatial and temporal resolution~\cite{Cheong2020,Nemec2018}.

\subsection{Birefringence imaging}
The first imaging of antiferromagnetic domains utilized birefringence imaging to visualize antiferromagnetic domains in the insulating collinear antiferromagnet NiO~\cite{Roth1958_2,Roth1960}. Similar to the Kerr or Faraday effect in ferromagnets, birefringence imaging is based on the polarization rotation of reflected or transmitted light from an antiferromagnetic material. While differences in the birefringence of antiferromagnetic domains can originate directly from second order magneto-optical effects (Cotton-Mouton or Voigt Effect)~\cite{Tzschaschel2017,Saidl2017}, birefringence of AFMs with a strong magnetoelastic coupling is often dominated by strain-induced birefringence~\cite{Dachs1973,Gebhardt1976, Gehring1977}. Birefringence imaging using a polarizing microscope is a powerful and easily accessible tool to investigate the antiferromagnetic domain structure of bulk crystals~\cite{Roth1960,Kondoh1964} and thin films~\cite{Xu2019a, Xu2020a}. Moreover, it can be readily combined with additional techniques such as current-induced switching experiments~\cite{Schreiber2020,Schreiber2021,Meer2021}, application of magnetic fields~\cite{Xu2022} or pump-probe techniques~\cite{Tzschaschel2017}. 

\subsection{X-ray magnetic dichroism}
\label{sec:XMLD}
Element specificity, sensitivity to chemical sites, and variable depth sensitivity make polarized X-ray absorption spectroscopy a powerful tool~\cite{Stohr1998}.
While X-ray magnetic circular dichroism (XMCD) vanishes for fully compensated magnetic moments, antiferromagnetic order can be studied by measuring the X-ray magnetic linear dichroism (XMLD)~\cite{Kuiper1993,Alders1998,Stohr1999,Scholl2000}. The XMLD signal is given by the difference in absorption of linearly polarized X-rays with polarization parallel and perpendicular to the N\'eel order (see Fig.~\ref{fig:Imaging}a).
In addition to XMLD, crystal fields can also induce linear dichroism, which has to be carefully disentangled  from the magnetic contribution~\cite{Arenholz2006}.

Today, XMLD detected by photoemission electron microscopy (PEEM) is one of the most widely used techniques to image antiferromagnetic domain structures. Vector maps reconstructed from angle-dependent X-ray imaging reveal the orientation of the N\'eel order~\cite{Moya2013,Chmiel2018}. In contrast to PEEM, scanning transmission X-ray microscopy (STXM) allows for measuring XMLD in sizable magnetic fields. The necessity to deposit the material of interest on a membrane can be circumvented by detecting the XMLD in the total electron yield (TEY)~\cite{Behyan2011,Wittmann2022}.

\subsection{Nitrogen-vacancy center magnetometry}
Nitrogen-vacancy (NV) centers in diamond have been shown to be highly sensitive and non-perturbative probes for sensing the stray field of magnetic materials, operating over a wide range of temperatures and magnetic fields, and a dynamic range spanning from direct current to gigahertz~\cite{Taylor2008,Rondin2014,Degen2017,Casola2018}.
While AFMs do not produce global stray fields, small magnetic stray fields can arise locally due to uncompensated magnetic moments at the surface, topographic features, or domain walls (see Fig.~\ref{fig:Imaging}b).
Wide-field NV microscopes use a camera to image a dense layer of NV centers close to the surface of a diamond crystal adjacent to the magnetic layer and are, hence, diffraction limited.
The spatial resolution of scanning NV magnetometry, which is based on scanning a single NV center at the apex of a diamond tip across the sample, can reach the nanometer scale.
A major challenge of NV magnetometry is the reconstruction of the magnetization from the stray fields. To lift the ambiguity of the reconstruction, a dual approach enabling detection of both, the stray field and the magnetic order, can be used~\cite{Lenz2021}.
Recent reports have successfully probed the magnetic domain structure of 3d metal oxide AFMs using NV magnetometry~\cite{Appel2019,Kosub2017a,Wornle2021,Welter2022} as well as the magnetic noise and spin waves via spin relaxometry~\cite{Du2017,Finco2021,Rollo2021,Wang2022}.

%%%%%%%%%%%%%%%%%%%%%%%%%%%%%%%%%%%%%%%%
\section{3d oxide materials and their spin structures}
In ferromagnets, magnetic domains are defined by regions of uniform magnetization. In order to minimize the stray field, ferromagnets typically adopt a multi-domain state in absence of an external magnetic field depending on the interplay of a range of relevant energy terms such as the exchange interaction, dipolar interaction, and anisotropy. While complex spin textures have been studied widely in ferromagnets, much of the underlying physics of their analogues in AFMs remains to be explored.

In collinear AFMs, the order parameter, the N\'eel vector, is given by the direction of the staggered magnetization of the sublattices. 
%Compared to conventional ferromagnets, where non-trivial spin structures forming domains are well understood to typically result from the interplay of the energy terms (dipolar interactions, anisotropy, exchange, etc.), the spin structures in AFMs are less well studied. While in ferromagnets the domain configuration is described by the local variation of the magnetic moment in space, in AFMs it is the local variation of the Néel vector (the difference of the magnetization directions in the two sublattices) that describes the domain structures. Given the zero net stray field of collinear AFMs, one would intuitively expect single domain states, where the Néel vector is pointing uniformly in a certain direction. Furthermore, g
Given the zero net magnetic moment, one might expect that magnetic fields are unable to change the N\'eel vectors. However, in order to explain the magnetic susceptibility observed in some antiferromagnetic materials, N\'eel proposed that domain walls separating different antiferromagnetic domains can be displaced by small magnetic fields~\cite{Neel1954}. Indeed, multi-domain states have been observed in a broad range of materials today~\cite{Roth1960,Nafradi2016,Scholl2000,Ince1968,Goltsev2003,Spooner1969,Zimmermann2009}.% For example the formation of domains in the antiferromagnet $\alpha$-Fe$_2$O$_3$ ($\alpha$-Fe$_2$O$_3$) was analyzed and it was found that the change in entropy due to the magnetic disorder in magnetic domains compared to the energy of the domain walls is far too small to account for the presence of domains with non-uniform Néel vector\cite{Li1956}.
The staggered magnetic order in collinear AFMs gives rise to a zero net stray field. Thus, based on the intuition developed for ferromagnets, spontaneous formation of domains in AFMs might come as a surprise. 
In antiferromagnetic oxides, strong magnetoelastic interactions come into play and can lead to the formation of domains, as discussed in more detail below. Antiferromagnetic domains resulting from magnetoelastic interactions predominantly form along different (noncollinear) directions~\cite{Gomonaj1999,Gomonay2002,Gomonay2018a,Wittmann2022}.
To understand the formation of 180$^{\circ}$ domains, we have to consider additional mechanisms~\cite{Rado1961,Mitsek1970}. For instance, 180$^{\circ}$ domains can result from defects where domains are nucleated during cooling from above the ordering temperature, the Néel temperature~$T_\mathrm{N}$~\cite{Appel2019}. Another possibility is the merging of two 90$^\circ$ domain walls when the antiferromagnet is driven above the spin flop transition~\cite{Mitsek1970}. During this process, two domains that have nucleated separately meet and may form a 180$^{\circ}$ domain wall. %The domain walls forming have a particular spin structures that is a consequence of the interplay between the magnetoelastic coupling, the exchange and anisotropies and can vary strongly between different systems as discussed below.
In AFMs, the formation of domains and complex spin textures is mainly governed by the interplay between exchange interaction, anisotropy, and magnetoelastic coupling and can vary strongly between different materials systems as discussed below.

\subsection{$\alpha$-Fe$_2$O$_3$}

A promising material candidate for antiferromagnetic spintronics is the well-known insulating AFM $\alpha$-Fe$_2$O$_3$ (hematite) in which recently long-distance transport of spin-information was observed (see section~\ref{sec:spintransport}). $\alpha$-Fe$_2$O$_3$ exhibits a range of exciting properties such as the Dzyaloshinskii-Moriya Interaction (DMI) leading to the Morin transition~\cite{Morrish1995}: at low temperatures the interplay between anisotropy and DMI favors an uniaxial anisotropy with the Néel vector oriented along the c-axis, while above the Morin transition (260$\,$K) the canting of the two sublattices favored by the DMI leads to a weak magnetic moment with the Néel vector perpendicular to the c-axis~\cite{Moriya1960}. In this canted moment state, the antiferromagnetic domains can be controlled by modest magnetic fields, leading to large single domain states~\cite{Bezencenet2011}. The Morin transition and the anisotropies can be tailored by doping and a strong dependence on the structure and the resulting direction of the easy axis has been observed~\cite{Ross2020,Jani2021}.

To study the magnetic properties, early experiments on bulk $\alpha$-Fe$_2$O$_3$ were carried out using neutron scattering~\cite{Curry1965,Besser1967}. Using electrical methods based on spin Hall magnetoresistance (see section~\ref{Sec:Electrical detection}), the anisotropies, DMI and the spin-flop have been determined as a function of temperature~\cite{Lebrun2019}. To understand the spin dynamics properties, additionally, antiferromagnetic resonance measurements~\cite{Lebrun2020} and spin pumping measurements \cite{Boventer2021} were carried out, and a low magnetic damping was established~\cite{Lebrun2020}.

\begin{figure}[ht]
    \centering
    \includegraphics[width = 0.4\textwidth]{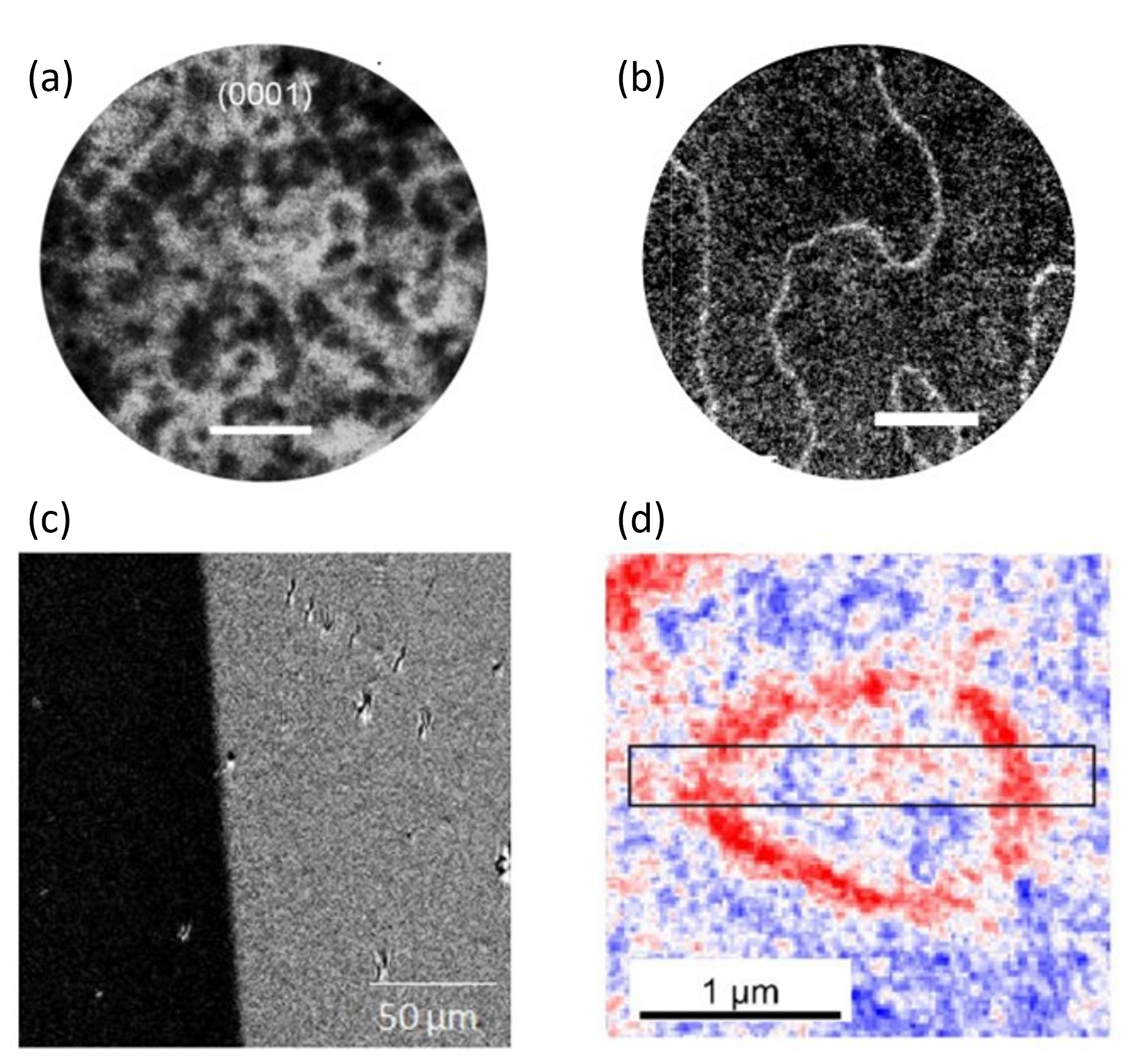}
    \caption{XMLD-PEEM images of the domain structures of (a) 100$\,$nm thick (0001) $\alpha$-Fe$_2$O$_3$ at 100$\,$K and (b) 100$\,$nm thick (1$\overline{1}$02) $\alpha$-Fe$_2$O$_3$ at 100$\,$K, the scale bar corresponds to 2$\,\mu$m. (c) Magneto-optical Kerr microscopy image of a bulk (1$\overline{1}$02) bulk crystal at room temperature showing large domains. (d) A spin structure with the topology of an antiferromagnetic antiskyrmion found in (1$\overline{1}$02) $\alpha$-Fe$_2$O$_3$. (a-b) Reprinted (adapted) with permission from Ref. \onlinecite{Ross2019a}. Copyright {2022} American Chemical Society. (d) from supplementary information of Ref.~\onlinecite{Ross2020}.}
    \label{fig:hematite}
\end{figure}

The spin structure in $\alpha$-Fe$_2$O$_3$ can be ascertained by magnetic imaging. In bulk $\alpha$-Fe$_2$O$_3$ crystals, large domains are observed, and macroscopically large areas of single domain can be found~\cite{Williams1958}. More interesting are thin films, where the growth-strain can generate additional domain configurations. As seen in Fig.~\ref{fig:hematite}, very different domain structures exist. In (0001) $\alpha$-Fe$_2$O$_3$ small domains are observed below the Morin transition (Fig.~\ref{fig:hematite} (a)), while in (1$\overline{1}$02) films, larger domains with clear domain walls are found (b). In the easy-plane phase above the Morin transition also multi-domain states are observed in thin films, while large domains can generally be stabilized in bulk crystals (Fig.~\ref{fig:hematite} (c)). The transition from the easy-axis phase to the easy-plane phase that occurs in the Morin transition also leads to a transition from 180$^\circ$ to 60$^\circ$ domain walls, given the six-fold in-plane symmetry. 

Finally, $\alpha$-Fe$_2$O$_3$ also exhibits topologically non-trivial spin structures (Fig.~\ref{fig:hematite} (d)), which is surprising since the symmetry of $\alpha$-Fe$_2$O$_3$ does not allow for Lifshitz-Invariants that could stabilize a chiral magnetic structure. The antiskyrmion-like structures observed in $\alpha$-Fe$_2$O$_3$ are likely rather domain structures stabilized by local pinning sites~\cite{Ross2020,Chmiel2018,Jani2021a}.

\subsection{Cr$_2$O$_3$}

\begin{figure}[hb]
    \centering
    \includegraphics[width = 0.4\textwidth]{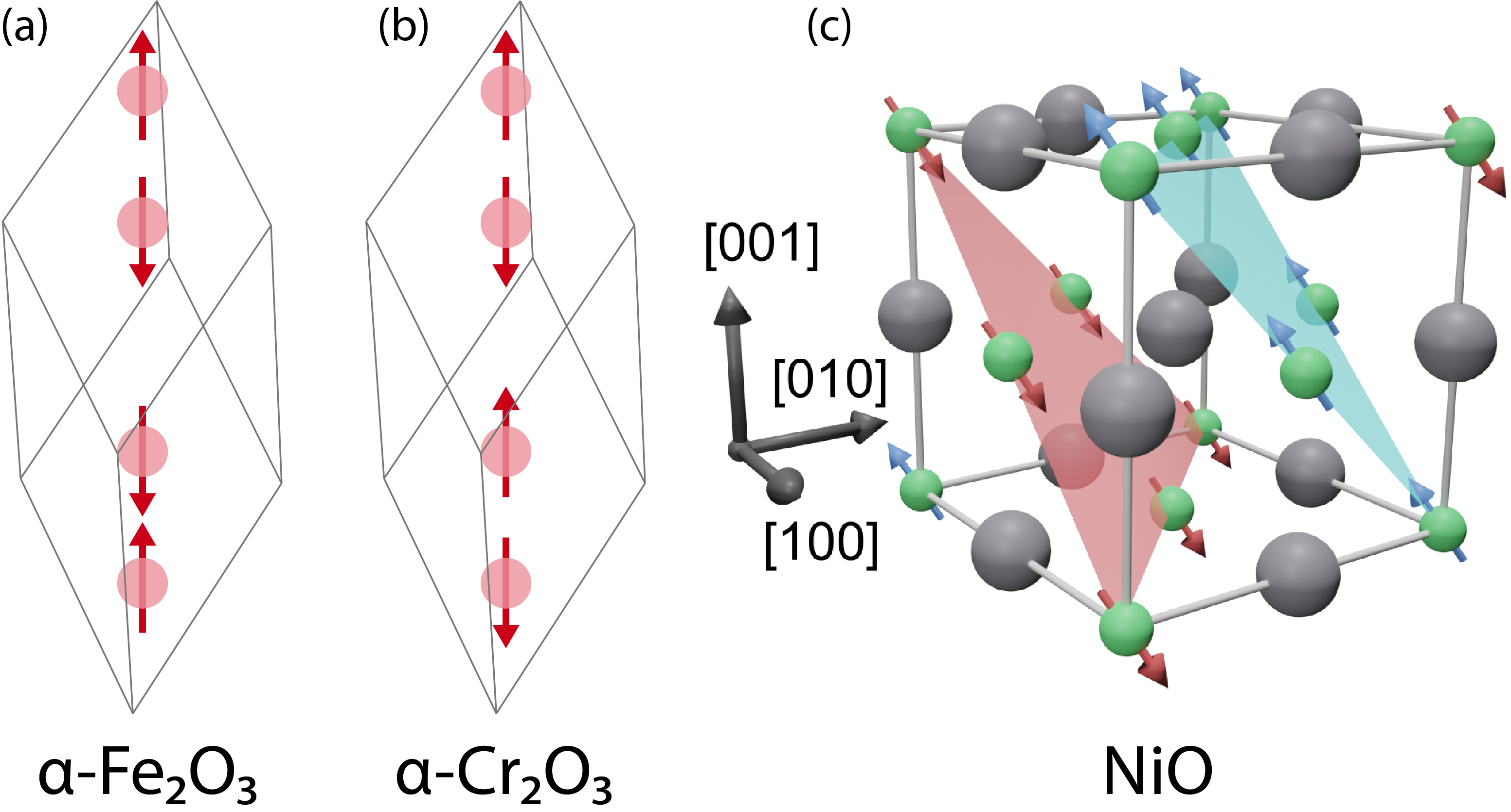}
    \caption{Spin structure of the AFMs (a)~$\alpha$-Fe$_2$O$_3$, (b)~$\alpha$-Cr$_2$O$_3$, and (c)~NiO.}
    \label{fig:UnitCell}
\end{figure}

Similar to $\alpha$-Fe$_2$O$_3$, the $\alpha$-Cr$_2$O$_3$~\cite{Brockhouse1953, Corliss1965} adopts a corundum structure and is an insulating collinear AFM below the N\'eel temperature ($T_\mathrm{N}=308\,$K~\cite{Foner1963}). The crucial difference between the two isomorphous $\alpha$-Fe$_2$O$_3$ and $\alpha$-Cr$_2$O$_3$ lies in the arrangement of the spins along the [111] axis as shown in Fig.~\ref{fig:UnitCell}~(a) and (b)~\cite{Dzyaloshinsky1958}. The symmetry of the magnetic crystal also gives rise to the linear magnetoelectric effect in Cr$_2$O$_3$~\cite{Dzyaloshinskii1959,Astrov1960,Rado1961}.

Due to the the easy-axis anisotropy in Cr$_2$O$_3$, there are two types of domains separated by 180$^\circ$ domain walls which makes the detection of the domain structure challenging. The first experimental observation of the domain structure was achieved via second-harmonic generation~\cite{Fiebig1995}.
Making use of the equilibrium boundary magnetization of magnetoelectric AFMs~\cite{Belashchenko2010}, the surface magnetization domains have been imaged by XMCD, magnetic force microscopy, and scanning NV magnetometry~\cite{Wu2011,Kosub2017a} revealing the existence of both Bloch and N\'eel type domain walls~\cite{Wornle2021}.
In addition to the magnetic imaging technique, the antiferromagnetic order has also been studied electrically in Cr$_2$O$_3$/ heavy metal heterostructures~\cite{Kosub2015,Schlitz2018, Ji2018}.

Recent work has demonstrated the high potential of Cr$_2$O$_3$-based devices for a wide variety of further research, ranging from spin superfluidity to novel memory technologies~\cite{Yuan2018,Kosub2017a,Hedrich2021}.

\subsection{NiO}

NiO was possibly the first material in which antiferromagnetic domains could be observed directly~\cite{Roth1958_2,Roth1960}. The high N\'eel temperature of T$_\text{N}=523\,$K~\cite{Henry1951} in the bulk makes NiO an attractive candidate for investigating antiferromagnetic domain structures and antiferromagnetic spintronics. 
Below its N\'eel temperature, NiO adopts a type-II antiferromagnetic order: the spins are antiferromagnetically coupled between the $\{111\}$ planes due to superexchange~\cite{Anderson1950} and ferromagnetically coupled inside the $\{111\}$ planes~\cite{Keffer1957} as shown in Fig.\ref{fig:UnitCell} (c). The exchange striction between the antiferromagnetically coupled planes leads to a contraction between the $\{111\}$ planes and a temperature dependent~\cite{Hutchings1972,Rooksby1948} rhombohedral distortion of the cubic crystal~\cite{Roth1960a,Shull1951,Smart1951,Smart1952}. The symmetry of the original fcc lattice allows four possible combinations of antiferromagnetically coupled $\{111\}$ planes, which are each associated with a different strain. These domains are considered to be Twin domains (T-domains)~\cite{Roth1960}. An additional smaller anisotropy due to dipolar interactions leads to a threefold set of preferential spin orientations within the ferromagnetic planes along the [112] directions, the so-called Spin domains (S-Domains)~\cite{Roth1960,Kondoh1964,Yamada1963,Yamada1966}. If one additionally considers $180^\circ$ domains, we arrive at a total of 24 possible domain configurations in the NiO bulk system.

The impact of strain on the antiferromagnetic domain structure has been studied extensively in NiO~\cite{Roth1960, Roth1958_2,Slack1960,Yamada1966,Yamada1966a,Yamada1966b,Kondoh1964,Baruchel1981,Kleemann1980,Weber2003a,Arai2012}. In particular, external strain such as strain during cleaving, polishing, sample glue, or handling with tweezers can easily manipulate the antiferromagnetic domains~\cite{STREET1951,Mandal2009,Kondoh1964,Kurosawa1974,Mandel1997}.
Furthermore, the domain structure of both, bulk and thin film NiO samples is influenced by the growth conditions such as growth temperatures~\cite{Giovanardi2003,Wu2008}, oxygen pressure~\cite{Myoungjae2004}, and film thickness~\cite{Alders1998}. Importantly, the strain resulting from the lattice mismatch with the substrate plays a significant role for the anisotropy~\cite{Finazzi2009,Wu2008}. %Compressive strain in NiO grown on MgO leads to an out-of-plane alignment~\cite{james1999,Alders1998} and tensile strain in NiO grown on Ag~\cite{Giovanardi2003} leads to a preferential in-plane alignment of the N\'eel vector~\cite{Altieri2003,Finazzi2003}.
Large domain sizes can be achieved by annealing~\cite{Roth1958_2,Saito1962}.
As a result, it is crucial to study the domain structure of NiO thin films by an imaging technique like XMLD-PEEM, to understand observations in current-induced switching experiments~\cite{Schmitt2020a,Schmitt2022}. 

\subsection{CoO}

First neutron diffraction data has indicated a similar magnetic structure for CoO as for NiO~\cite{Shull1951}. However, in contrast to NiO, the orbital moment in CoO is not quenched and CoO exhibits a strong spin-orbit coupling, which affects the orientation of the spins~\cite{Shull1951,Kanamori1957,Ghiringhelli2002}.
Investigations of the crystallographic structure of CoO revealed a tetragonal distortion of the fcc lattice below the N\'eel temperature (290$^\circ\,$K) in contrast to the rhombohedral distortion in NiO or MnO~\cite{TOMBS1950,Greenwald1950,Li1955}.
Neutron diffraction measurements could not provide clear information on the spin structure of CoO and different models were proposed~\cite{Jauch2001}. Initially, a collinear spin structure was assumed in which the spins are ferromagnetically coupled inside the $\{111\}$ planes~\cite{Shull1951,Roth1958,nagamiya1958,VanLaar1965}. However, the tetragonal symmetry of the crystal led to the proposal of a noncollinear spin structure in a multi-spin-axis model~\cite{VanLaar1965,VanLaar1965a,VanLaar1966}.
Further studies revealed that the tetragonal deformation can occur in a collinear spin alignment and is additionally accompanied by a monoclinic deformation~\cite{Kanamori1957,Kanamori1957a,Saito1966}. After the experimental observation of the additional monoclinic distortion, the magnetic structure of CoO is widely accepted to be collinear~\cite{Nagamiya1965,Saito1966,Herrmann-Ronzaud1978,Germann1974a,Saito1966,Jauch2001,Nakanishi1974}. However, we note that recent studies have again proposed a noncollinear structure for CoO~\cite{Tomiyasu2004,Kruger2021,Kruger2022}. 
In the collinear structure, the spins of CoO are coupled ferromagnetically inside the $\{111\}$ planes and antiferromagnetically between the planes. The tetragonal distortion of CoO along one of the three cubic axes leads to the possible formation of three different twin domains, compared to four different twin domains in NiO~\cite{Spooner1969}
Large domains in CoO crystals can be obtained by annealing the crystals above the N\'eel temperature~\cite{Uchida1964,Herrmann-Ronzaud1978,Saito1966,Rechtin1971} and the application of external strain can lead to the preferential stabilization of certain domains~\cite{Herrmann-Ronzaud1978}. %The magnetoelastic coupling can be observed in birefringence measurements~\cite{Germann1974a} and can be utilized to image the domains of CoO crystals in x-ray diffraction measurements~\cite{Saito1966}.
Similar to NiO, the magnetic structure of CoO depends on the growth parameters and on the substrate-induced strain~\cite{Spooner1969,Li2015}. In thin films, compressive strain favors an in-plane alignment of the magnetic moments~\cite{Csiszar2005,Zhu2014a} and tensile strain leads to an out-of-plane alignment~\cite{Csiszar2005,Zhu2014a}. For CoO grown on MgO(001) the spins align in the plane of the film~\cite{Zhu2014} and two in-plane magnetic easy axes along [$110$] or [$\bar{1}10$] are present~\cite{Baldrati2020,Cao2011}. Birefringence imaging can be used to image the domains of thin-films~\cite{Xu2020a,Xu2022} and second-order magneto-optical effects can also be used to study the magnetization dynamics of CoO thin films~\cite{Zheng2018}. 
CoO is often considered to be similar to NiO, however, the unqueched orbital moment leads to significant differences in its crystallographic structure, magnetic properties~\cite{Grzybowski2021a}, and rich electrical properties~\cite{Sarte2020}.

%%%%%%%%%%%%%%%%%%%%%%%%%%%%%%%%%%%%%%%%%
\section{Antiferromagnetic Shape Anisotropy}

\begin{figure}[h]
    \centering
    \includegraphics[width = 0.4\textwidth]{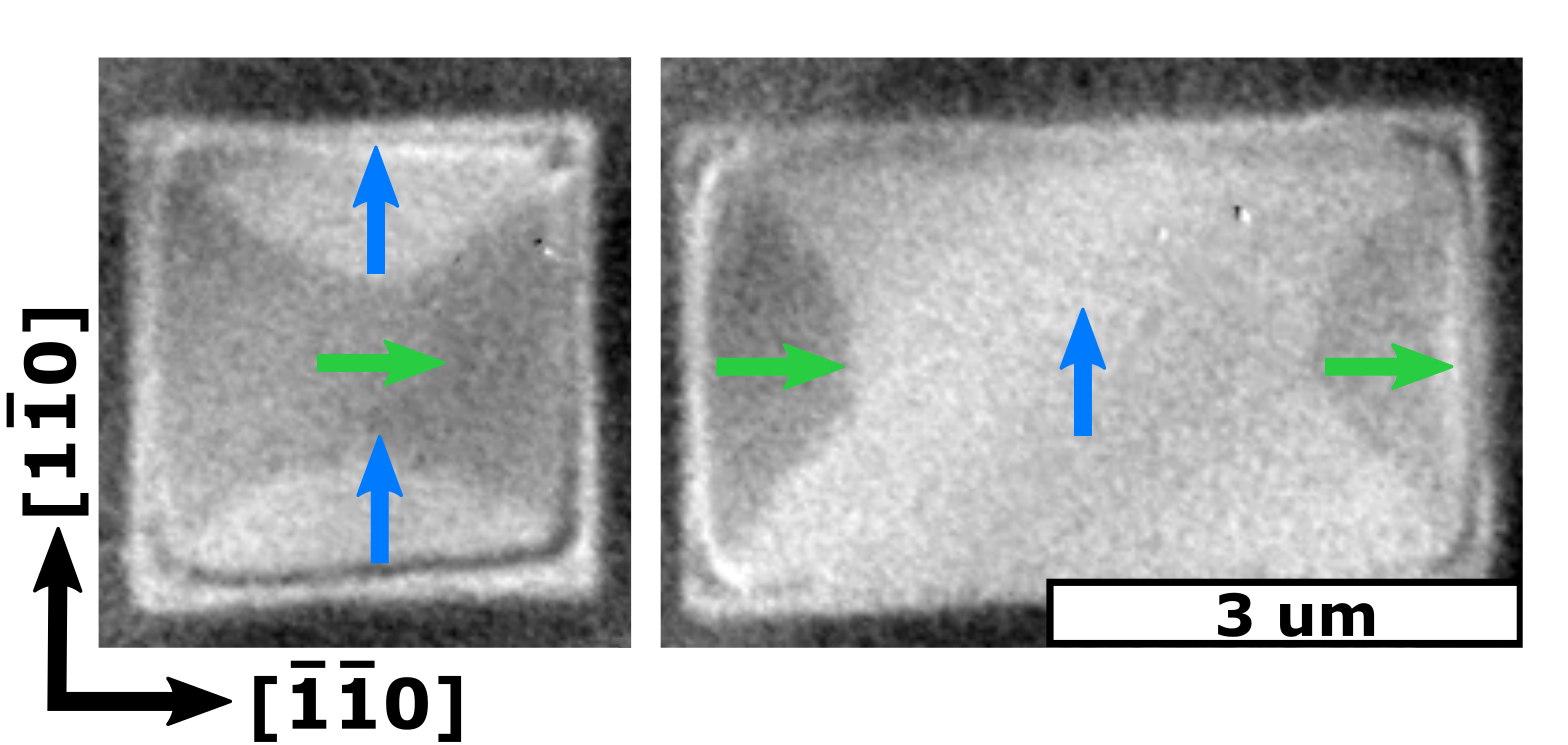}
    \caption{ XMLD-PEEM image of the domain structure in  patterned NiO/Pt bilayer devices after annealing. The arrows indicate the direction of the inplane projection of the N\'eel vector \cite{Meer2022}.}
    \label{fig:NiOShape}
\end{figure}

In ferromagnets shape anisotropy due to stray fields has long been a key tool to tailor device properties. In AFMs this effect is absent due to the lack of a net magnetic moment.
Nevertheless, it has been predicted that in AFMs with a strong magnetoelastic coupling, strains can lead to antiferromagnetic shape-anisotropy~\cite{Gomonay2007,Gomonay2014}. First investigations of this effect were carried out on antiferromagnetic LaFeO$_3$ and local changes of the domain structure near the edge of patterned devices were observed~\cite{Folven2010,Folven2011,Folven2012,Folven2012a}. 
However, recent investigations on NiO have revealed that the patterning of devices can lead to two different effects. In devices patterned along the projection of the easy axes of NiO, local changes of the surface anisotropy lead to a preferential stabilization of certain domains in the vicinity of the patterned edge. In addition to this, long-range magnetoelastic strains which originate from the domain formation determine the equilibrium domain structure in the center of the device. The strain during the formation of the individual domains determines the domain structure and can be used to control the antiferromagnetic ground state, see Fig.~\ref{fig:NiOShape}~\cite{Meer2022}. 
Therefore, antiferromagnetic shape anisotropy is a powerful tool to engineer antiferromagnetic domain structure in devices and is not limited to NiO, but can be extended to other AFMs with strong magnetoelastic coupling such as CoO or $\alpha$-Fe$_2$O$_3$.

 %%%%%%%%%%%%%%%%%%%%%
% Current-induced effects
%%%%%%%%%%%%%%%%%%%%%%

\section{Current-induced effects}
\subsection{Electrical detection}\label{Sec:Electrical detection}
Electrical detection of the antiferromagnetic order is one of the main building blocks for realizing antiferromagnet-based memory technologies. For insulating AFMs, electrical reading of the magnetic state is commonly achieved in antiferromagnet/ heavy-metal bilayer structures making use of the spin Hall magnetoresistance (SMR)~\cite{Chen2013a,Manchon2017}.
A charge current in a heavy-metal creates a spin current flowing toward the interfaces due to the spin Hall effect. The interfacial interaction of the spins in the heavy-metal with the magnetic insulator depends on the relative orientation of the spins and the magnetic order. The interfacial exchange of angular momentum can be detected electrically via the inverse spin Hall effect in the heavy-metal. This gives rise to the characteristic $\sin(2\theta)$ dependence of the SMR on the angle~$\theta$ between the charge current direction and the magnetic order.
In contrast to ferromagnets, the perpendicular alignment of the N\'eel vector with respect to the external magnetic field results in a negative sign of the SMR signal for AFMs~\cite{Han2014,Hoogeboom2017,Fischer2018,Lebrun2019,Geprags2020b}.
Given the simple planar device structure required for electrical detection, SMR has proven to be a powerful tool for probing magnetic properties and enables the integration of AFMs in the next-generation computing technologies.

\subsection{Spin Transport}
\label{sec:spintransport}

\begin{figure}[ht]
    \centering
    \includegraphics[width = 0.47\textwidth]{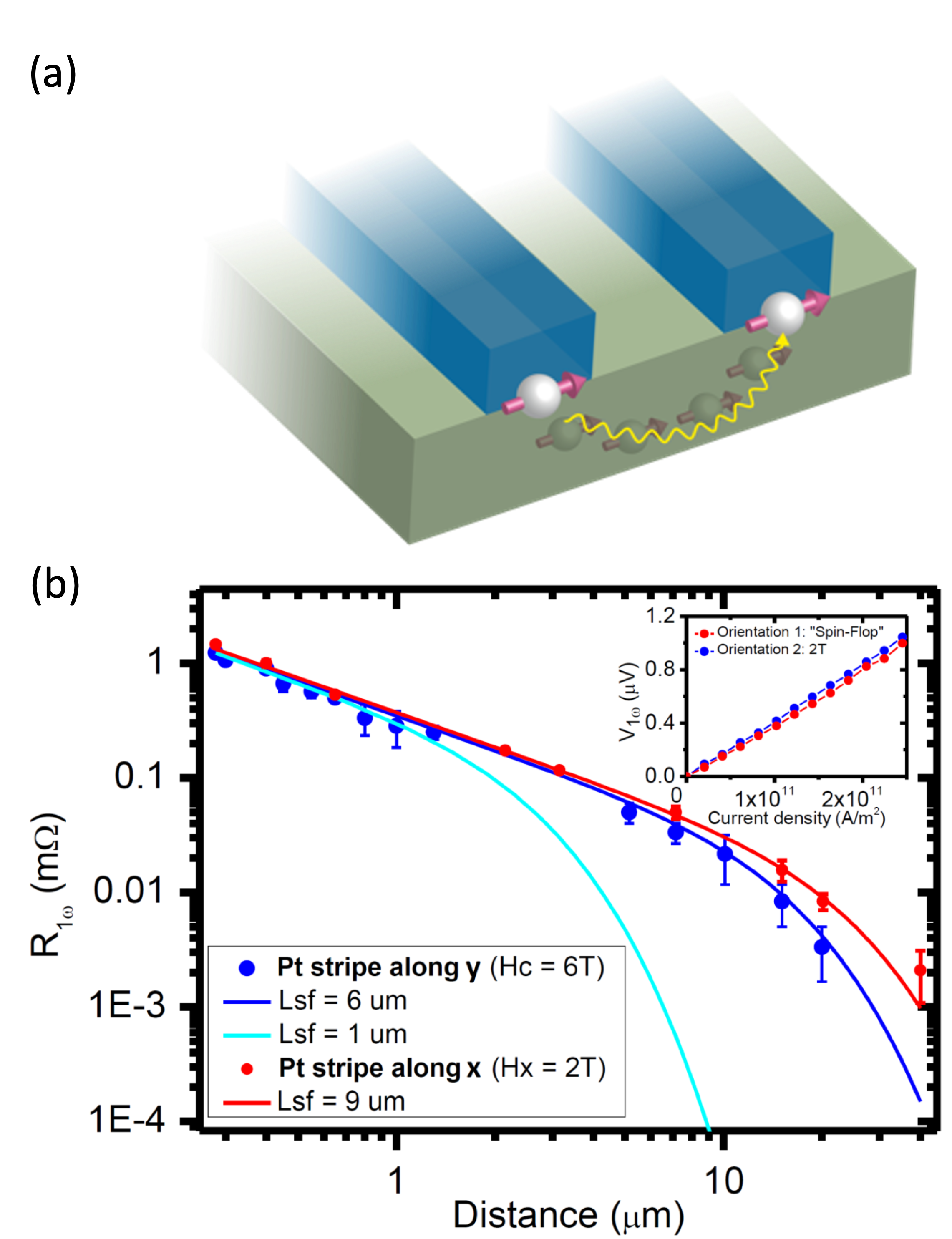}
    \caption{(a) Schematic of non-local spin transport. Spins are injected by the spin Hall effect from a heavy metal wire (left blue injection wire) as a magnonic spin current into the antiferromagnetic insulator (gray), where they propagate to a second heavy metal wire (right blue detection wire). There, the spin current is absorbed and detected electrically by the inverse spin Hall effect. (b) Measured non-local resistance as a function of distance in a bulk $\alpha$-Fe$_2$O$_3$ (1$\overline{1}$20) crystal, showing spin transport over micrometer distances. The inset shows a linear dependence of the signal on the current density with no threshold, indicating diffusive transport. (b) from Ref.~\onlinecite{Leb2018}.}
    \label{fig:transport}
\end{figure}

Many new ideas regarding the generation and transport of spin-currents in antiferromagnetic materials have been generated in the last few years, but most of them remained at the stage of predictions. Conventional experimental work has so far mainly consisted of studying AFMs in coupled AFM/FM systems~\cite{Zhang2014,Wang2014,Hahn2014,Frangou2016} in which the antiferromagnetic properties are deduced from a detection in the adjacent FM, as for instance realized in a magnon spin-valve device~\cite{Cra2018}. The transport perpendicular to the plane across the layer thickness in antiferromagnetic insulators with adjacent ferro-(ferri-)magnetic layers has typically yielded only very short spin transport length scales of a few to a few tens of nm~\cite{Bal2018, Wang2014, Lin2016}. To study the pure AFM transport without other magnetic layers involved, non-local transport (see Fig.~\Ref{fig:transport} (a)) is a suitable technique. Recently, some of the longest known spin transport length scales were shown for low damping antiferromagnetic insulators like YFeO$_3$~\cite{Das2022}, Cr$_2$O$_3$ \cite{Yuan2018} and $\alpha$-Fe$_2$O$_3$ in the easy axis phase (Fig.~\Ref{fig:transport} (b))~\cite{Leb2018}.
While spin transport is expected in the easy-axis phase due to the circular polarization of the magnons, lateral spin transport has also been demonstrated in the easy-plane phase of $\alpha$-Fe$_2$O$_3$. Fundamentally, no magnonic spin transport is expected because the magnons in easy-plane AFMs are in the simplest description linearly polarized. However, pairs of linearly polarized magnons with different k-vectors but the same energy can be combined into circularly polarized magnons, leading to magnons with a certain degree of circular polarization and thus transport of spin angular momentum~\cite{Lebrun2020, Han2020, Wim2020}. This means that in easy-plane and orthorhombic AFMs, one can electrically generate pairs of linearly polarized spin waves, which carry an effective circular polarization and thus spin information~\cite{Ross2020}.
  
%Tuning the interplay between the DMI leading to the weak moment of $\alpha$-Fe$_2$O$_3$ above the Morin transition and a hysteresis in the field dependent non-local long distance spin transport. This shows that one can obtain bi-stability of states, allowing non-volatile switching between spin transport regimes~\cite{Ross2022}.
%To understand the origin of the long distance transport, we have carried out spin dynamics measurements of both the high frequency spin dynamics modes up to hundreds of GHz frequencies \cite{Lebrun2020} as well as the low frequency mode \cite{Boventer2021}. In particular, we determined the damping in $\alpha$-Fe$_2$O$_3$ to be $10^{-5}$. Thus, making this material one of the lowest damping materials available, on a par with the ferrimagnetic insulator YIG.

Finally, by linking the spin structures to the transport, it has been found that the domain structure signinficantly impacts the  length scale of spin transport. The domain walls act as efficient spin scatterers and thus strongly governing the spin transport~\cite{Ross2020,Ross2019a}.

\subsection{Electrical manipulation}
In addition to reading, writing AFMs efficiently is an actual key open goal of the field. One possible approach relies on staggered Néel spin-orbit torques, creating a staggered effective field of opposite sign on each magnetic sublattice for special crystallographic compounds, as studied in the particular metallic compounds CuMnAs and Mn$_2$Au~\cite{Wadley2016a,Bodnar2018}. An approach, applicable to a broader range of AFMs including the 3d oxide materials reviewed here, is to use the non-staggered, antidamping-like torque exerted by a spin accumulation at the interface of a heavy metal and an AFM-insulator due to a charge current in the heavy metal~\cite{Shiino2016,Gomonay2016}. This mechanism leads to a displacement of domain walls that can be potentially very fast due to the absence of a Walker breakdown phenomenon~\cite{Gomonay2016}. Here, the direction of the wallmotion is governed by their sense of rotation. If there are no Lifshitz - invariants that break the symmetry and lead to a chiral spiralization of the AFM domain walls, this means that different domain walls move in different directions and no overall switching is achieved. In particular, with the recent observation that domain walls in NiO/Pt are not chiral~\cite{Schmitt2022}, this mechanism would be difficult to use for devices. Another mechanism is a ponderomotive force effect that has been predicted to yield deterministic 90$^\circ$ domain wall motion \cite{Baldrati2019}.

Experimentally, the possibility of switching was demonstrated in NiO/Pt~\cite{Chen2018,Moriyama2018, Baldrati2019,Gray2019, Schreiber2020} and also in heterostructures of CoO~\cite{Baldrati2020} and $\alpha$-Fe$_2$O$_3$~\cite{Cheng2020,Cogulu2021}.
As contradicting directions of the switching were reported, other effects beyond spin orbit torques had to be invoked. 
In particular, by comparing direct imaging and transport data, it was found that some of the transport-based detection suffered from artefacts due to electromigration. Resistances changed due to structural changes of the material, rather than switching of the magnetization~\cite{Baldrati2020,Matalla-Wagner2019,Chiang2019,Churikova2020, Schreiber2020}.

\begin{figure}[h]
    \centering
    \includegraphics[width = 0.45\textwidth]{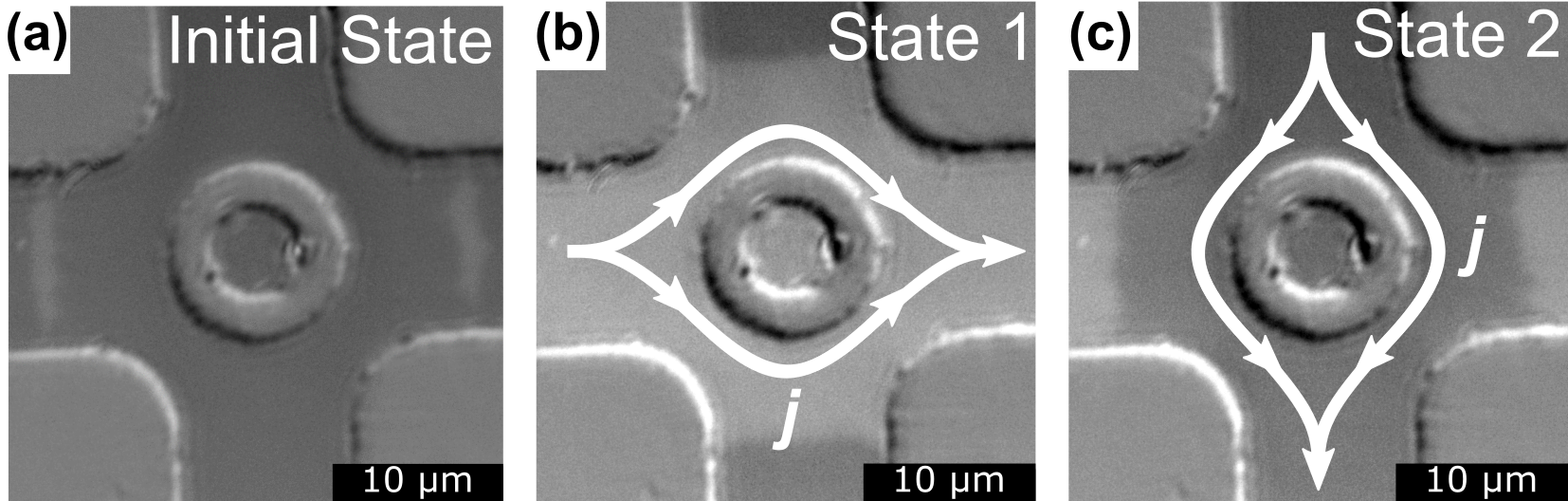}
    \caption{ (a) Birefringence image of the domain structure in a NiO/Pt bilayer device used for current-induced switching experiments before any current application. The circular area in the center of the cross is electrically isolated from the rest of the device. (b) When a current is applied to the device, the current flows around the dot. Nevertheless, after the current pulse has been applied, the domain inside the electrically isolated part is switched. (c) The application of an orthogonal pulse leads to orthogonal heat-induced strain and the dot can be switched back. Reprinted (adapted) with permission from Ref. \onlinecite{Meer2021}. Copyright {2022} American Chemical Society. }
    \label{fig:NiOSwitching}
\end{figure}

Furthermore, the magnetic switching observed by direct imaging also turned out to comprise contradicting directions of the switching, indicating that not only spin-orbit torques are involved. In a careful analysis, it was shown that localized heating leading to spatially varying strain induces switching by magneto-elastic coupling that depends strongly on the device geometry~\cite{Meer2021,Baldrati2020,Zhang2019}. As shown in Fig.~\Ref{fig:NiOSwitching}, switching can even be achieved in areas where no current is flowing, indicating that the governing mechanism in NiO/Pt bilayers studied here is not due to spin-orbit torques.

%%%%%%%%%%%%%%%%%%%%%
% Light-induced effects
%%%%%%%%%%%%%%%%%%%%%%

\section{Light-induced effects}
While the current-induced torques enable switching between different equilibrium antiferromagnetic states, the fast intrinsic dynamics of AFMs can be accessed effectively by optical pulses. One of the most promising applications of AFM-based devices is the optical generation of THz radiation induced by oscillations of the N\'eel vector.

\begin{figure}[h]
    \centering
    \includegraphics[width = 0.45\textwidth]{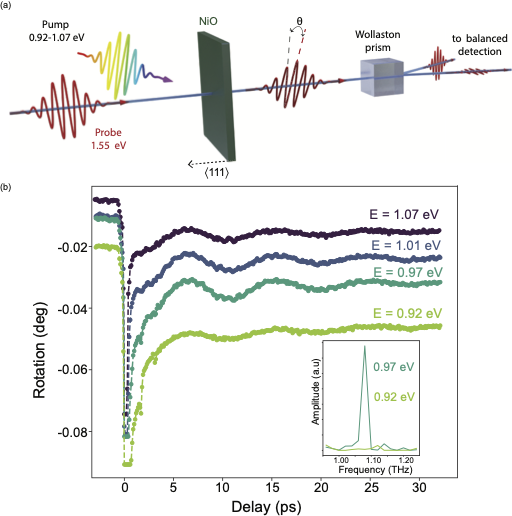}
    \caption{ (a) Typical setup for optical pump-probe experiments. (b)~Selected pump-probe traces for different pump-photon energies. Oscillations are associated with high-frequency (~1 THz) magnon modes in the NiO crystal. The inset shows power spectra of the signal, displaying the presence and absence of the 1~THz magnon mode (high peak) and domain wall oscillations (small peak).  Reprinted (adapted) with permission from Ref. \onlinecite{Bossini2021}. Copyright {2021} American Physical Society. }
    \label{Fig_light_induced}
\end{figure}

Light-induced excitation of the magnetic dynamics can be governed by different mechanisms. Direct excitation via magneto-optical effects (inverse Faraday effect, Cotton-Mouton effect) induces coherent rotation of the N\'eel vectors.  It occurs on short (subpicosecond) time scales and is sensitive to the polarization of the light. According to first-principle calculations~\cite{Lefkidis2007} and atomistic simulations~\cite{Dannegger2021}, the polarization should ideally be circular. Experimentally, optically-induced coherent magnetic dynamics and generation of THz magnons were observed in NiO~\cite{Duong2004,Fiebig2008, Satoh2010a,Nishitani2012,Qiu2021} and in Cr$_2$O$_3$~\cite{Duong2004,Fiebig2008}. However, magnon dynamics were also induced by linearly polarized pulses~\cite{Nishitani2012}, which was attributed to the multi-domain structure of NiO~\cite{Higuchi2016}. A less explored pathway to light-induced generation of coherent magnons involves excition-magnon transitions. Using this technique, the ultrafast spin dynamics of multi-domain NiO was explored~\cite{Bossini2021} (see Fig.~\ref{Fig_light_induced}).

Another mechanism of light-induced magnetic dynamics in AFMs is associated with the direct coupling of light with the electronic orbital angular momentum of the magnetic atoms, as demonstrated in a CoO single crystal~\cite{Satoh2017a}. Direct energy exchange between hot electrons of a heavy-metal (Pt) and the localized magnetic moments of an antiferromagnet (NiO) can result in optically-induced ultrafast reduction of the sublattice magnetization~\cite{Wust2022}.

In insulating antiferromagnet/ heavy-metal multilayers, laser-induced heating dominates over magneto-optical effects. In particular, laser-induced dynamics observed in NiO/Pt films~\cite{Rongione2022} were explained by the spin Seebeck effect combined with the ultrafast thermo-magnetoelastic mechanism. Both effects appear due to the temperature gradient between Pt and NiO layers, that induces shock strain waves and spin torques acting on the antiferromagnetic layer. The heating-induced response is slower compared to the magneto-optical response, but is insensitive to light polarization.
In addition to dynamic manipulation of the magnetic order, recent observations report that optically-induced heating can also lead to a polarization-independent creation of antiferromagnetic domains \cite{Meer2022a}.

All-optical manipulation of antiferromagnetic state was demonstrated both in single crystals~\cite{Duong2004,Fiebig2008, Satoh2010a,Satoh2017a,Bossini2021,Stremoukhov2022} and thin films with a heavy metal layer~\cite{Nishitani2012,Qiu2021,Rongione2022,Wust2022}. In the latter case, the heavy metal enables magnon-to-light conversion via inverse spin Hall effect and re-emission of optical pulses.

Both the electrical and optical manipulation of antiferromagnetic states involve magnetic dynamics that can be excited either by spin torques or by varying the magnetic anisotropy landscape. Spin torques can be excited in insulators, by spin pumping via spin Hall effect~\cite{Lebrun2019,Lebrun2020,Boventer2021}, by spin-transfer torque from optically induced spin polarized photoelectrons~\cite{Chirac2020}, or via the inverse Faraday effect~\cite{Satoh2010a}. The first two effects are interfacial and are efficient in thin films. In 3d metal oxides with strong magnetoelastic coupling, varying the magnetic anisotropy by stress or strain is also an efficient tool for spin switching. Similarly to current-induced switching, laser-induced heating of heterostructures creates additional shear strains that set preferable orientation of the magnetic vectors and induce switching. Remarkably, in addition to quasi static switching~\cite{Meer2021}, such thermo-magnetoelastic effects in NiO/Pt bilayers were also observed in the THz dynamics~\cite{Rongione2022}. Nevertheless, in real devices, both switching mechanisms, spin-assisted (by spin torques) and non-spin-assisted (by strain), coexist. Proper tailoring of the sample geometry can optimize the combination of both mechanisms and, thus, the switching. 

\section{Outlook}

Insulating 3d metal oxides are a promising class of materials for future antiferromagnetic devices, offering a variety of approaches for controlling and reading the antiferromagnetic order. A key advantage over their metallic counterparts is the observed long distance spin transport in these antiferromagnetic insulators~\cite{Leb2018,Das2022}. Transporting information without transporting charge carriers is a key aspect of insulating spintronics and builds the foundation for future energy-efficient spintronic devices. 

Due to the pronounced magnetoelastic coupling in AFMs, the domain structure is very sensitive to strain. As a result, strain can be used to tailor device properties during the growth or by strain-induced shape anisotropy. This static approach known as “straintronics” allows for the design of energy-efficient devices~\cite{Pellegrino2009,Bukharaev2018a}.
Dynamic manipulation of the N\'eel order via current-induced switching is based on two competing mechanisms, spin-orbit torque-based switching and switching due to current-induced heat and strain.
Optically induced switching and detection of the antiferromagnetic order extends the study of insulating transition metal oxides into the ultrafast regime, making them an attractive candidate for THz emitters due to their resonance frequencies in the THz range~\cite{Kampfrath2011}.

Another area in which antiferromagnetic 3d metal oxides could play a crucial role is the emerging field of “orbitronics”, where the orbital angular momentum is used to control the magnetic ordering~\cite{Go2021}. Comparing 3d metal oxides with a quenched orbital moment, such as NiO, and strong orbital moment, like CoO, could be used to study the influence of the orbital moment.

In recent years, several key concepts for the control of the antiferromagnetic ordering in 3d metal oxides have been developed. Today, a major challenge is the exploitation of the combination of these fundamental phenomena in spintronic devices.
Overall, the extensive research into the fundamental properties of transition metal oxides and the recent advances in the reading and writing of their antiferromagnetic ordering in thin films make them promising candidates for the exploration of novel concepts in future antiferromagnetic spintronic devices.

%\input{05_Discussion}
%%%%%%%%%%%%%%%%%%%%%%%%%%%%%%%Apendix+Notes for Appendix%%%%%%%%%%%%%%%%%%%%%%%%%%%%%%%%%%
\begin{acknowledgments}%preliminary

%%%%%%%%%%%%%
% Acknowledgements
M.K. thanks the IEEE Magnetics Society for support as a Distinguished Lecturer. Discussion with colleagues during the Distinguished Lecturer Tour have allowed many of the ideas and concepts presented here to be to be refined and further developed. 

M.K. acknowledge financial support from the Horizon 2020 Framework Programme of the European Commission under FETOpen Grant Agreement No. 863155 (s-Nebula).  All authors from Mainz also acknowledge support from MaHoJeRo (DAAD Spintronics network, Project Nos. 57334897 and 57524834) and KAUST (No. OSR-2019-CRG8-4048.2). The work
including the Mainz-Trondheim collaboration was additionally supported by the Research Council of Norway through its Centres of Excellence funding scheme, Project No. 262633 “QuSpin.”

The authors thank the German Research Foundation (SFB TRR 173 Spin+X 268565370 projects A01, B02,A11,B12,B14; TRR 288 – 422213477 (project A09) and project DFG No. 423441604), O.G.additionally acknowledges support from the ERC Synergy Grant SC2 (No. 610115), EU FET Open RIA Grant no. 766566
We thank A. R9oss for helpful discussions.

\end{acknowledgments}
%\appendix

%\input{Supplementary}

\bibliography{AFMreview}% Produces the 5liography via BibTeX.apssamp

\end{document}